\documentclass[prb,twocolumn,superscriptaddress,preprintnumbers,a4paper,amsmath,amssymb,showpacs,floatfix]{revtex4}
\usepackage{graphicx}
\usepackage{bm}

\begin{document}

\newcommand{\inA}{\mbox{\AA$^{-1}$}}
\newcommand{\q}{$\delta$}
\newcommand{\qlMn}{$q_{l}^{\rm Mn}$}
\newcommand{\qlDy}{$q_{l}^{\rm Dy}$}
\newcommand{\qlTb}{$q_{l}^{\rm Tb}$}
\newcommand{\qmMn}{$q_{m}^{\rm Mn}$}
\newcommand{\qmDy}{$q_{m}^{\rm Dy}$}
\newcommand{\qmTb}{$q_{m}^{\rm Tb}$}
\newcommand{\qMn}{$q^{\rm Mn}$}
\newcommand{\qmn}{$\delta_{m}^{\rm Mn}$} 
\newcommand{\qdy}{$q^{Dy}$} 
\newcommand{\dmn}{$\delta^{\rm Mn}$} 
\newcommand{\ddy}{$\delta^{\rm Dy}$} 
\newcommand{\bb}{$\mathbf{b}^{*}$}
\newcommand{\qc}{$\delta_{m}=\frac{1}{4}$} 
\newcommand{\qcs}{$\delta_{l}=\frac{1}{2}$} 
\newcommand{\TNMn}{$T_{N}$} 
\newcommand{\TNDy}{$T_{N}^{\rm Dy}$}
\newcommand{\TNTb}{$T_{N}^{\rm Tb}$} 
\newcommand{\Tn}{$T_{N}$} 
\newcommand{\TN}{$T_{N}$} 
\newcommand{\Tf}{$T_{f}$} 
\newcommand{\etal}{\textit{et al.}} 
\newcommand{\degg}{$^{\circ}$} 
\newcommand{\pa}{$P\|a$} 
\newcommand{\pb}{$P\|b$} 
\newcommand{\pc}{$P\|c$}


\title{Commensurate Dy magnetic ordering associated with incommensurate lattice distortion in orthorhombic DyMnO$_{3}$}
\author{R.~Feyerherm}
\affiliation{Hahn-Meitner-Institut (HMI), c/o BESSY, D-12489 Berlin, Germany}

\author{E.~Dudzik} 
\affiliation{Hahn-Meitner-Institut (HMI), c/o BESSY, D-12489 Berlin, Germany} 

\author{N.~Aliouane} 
\affiliation{Hahn-Meitner-Institut (HMI), Glienicker Str.~100, D-14109 Berlin, Germany} 

\author{D.~N.~Argyriou} 
\affiliation{Hahn-Meitner-Institut (HMI), Glienicker Str.~100, D-14109 Berlin, Germany} 

\date{\today}
\begin{abstract}
Synchrotron x-ray diffraction and resonant magnetic scattering experiments on single crystal DyMnO$_3$ have been carried out between 4 and 40~K. Below \TNDy = 5~K, the Dy magnetic moments order in a commensurate structure with propagation vector 0.5~\bb. Simultaneous with the Dy magnetic ordering, an incommensurate lattice modulation with propagation vector 0.905~\bb\space evolves while the original Mn induced modulation is suppressed and shifts from 0.78~\bb\space to 0.81~\bb. This points to a strong interference of Mn and Dy induced structural distortions in DyMnO$_3$ besides a magnetic coupling between the Mn and Dy magnetic moments.

\end{abstract}

\maketitle

Magnetic ferroelectrics, also referred to as \emph{multiferroics}, are materials where  magnetism and ferroelectricity are strongly coupled. They hold the promise for applications in nonvolatile data storage where information may be recorded by controlling the direction of the electric polarization by an applied magnetic field. The recent revival of interest in these materials is owed to the discovery of new frustrated spin systems that exhibit multiferroic behavior.\cite{fiebig:review} Recent highlights in this field are the observation of coupled magnetic and electric domains,\cite{fiebig} the finding that in some compounds either the ferroelectric polarization can be controlled magnetically\cite{kimura} or the magnetic phase can be changed by an electric field,\cite{lotter} as well as experimental and theoretical advances regarding the coupling of spiral magnetism to the lattice.\cite{kenz,katsura:057205,mostovoy:067601} 

In the orthorhombic rare earth (RE) manganites REMnO$_{3}$ (RE = Gd, Tb, and Dy), which crystallize in a distorted perovskite structure, ferroelectricity has been shown to originate from a spiral magnetic structure that breaks both time-reversal and inversion symmetry.\cite{mostovoy:067601,kenz} Here, the magnetic ordering of the Mn moments with incommensurate propagation vector \qmMn \bb\space is associated with a second-harmonic lattice modulation \qlMn = 2\qmMn, i.e., with a quadratic magnetoelastic coupling.\cite{PW,walker} On cooling below \TN (= 43, 41, and 39~K for RE = Gd, Tb, and Dy, respectively), the incommensurate values of \qMn\space vary with temperature and below a characteristic temperature \Tf\space ($\sim$ 23, 28, 18~K, respectively), a spontaneous electric polarization is observed in zero applied magnetic field for RE = Tb and Dy,\cite{kimura} while in GdMnO$_{3}$ ferroelectricity only occurs for applied magnetic fields $> 1$~T.\cite{kimura3}

For all three compounds, magnetic ordering of the RE magnetic moment has been reported to occur below $\sim$ 10~K.\cite{kimura3} Only for TbMnO$_{3}$ the magnetically ordered structures have been investigated in detail by neutron diffraction\cite{kenz,Quezel,kajimoto} a technique not well applicable to Dy and Gd compounds because of their very large neutron absorption cross sections. In TbMnO$_{3}$, the value of \qmMn\space varies between $\sim$ 0.28...0.29. Below \TNTb $\sim$ 7~K, Tb magnetic ordering takes place with propagation vector \qmTb \bb, \qmTb $\sim 3/7$. A significant magnetoelastic coupling is present for the Tb as indicated by the presence of a second-harmonic modulation 2\qmTb \bb\space below \TNTb\space in neutron diffraction data.\cite{kajimoto} 
The RE spins appear to be important for the magnetic control of polarization in the orthorhombic manganites. For example Kimura \etal\cite{kimura,kimura3} suggest that the field induced polarization flop in TbMnO$_{3}$ from \pc\space to \pa\space is driven in part by a metamagnetic transition of the Tb spins. Recent theoretical modeling of the polarization flop using Landau theory fails to predict the correct behavior under magnetic field in the same material unless a coupling between Mn and highly anisotropic Tb spins is considered.\cite{mostovoy:067601} 

For DyMnO$_{3}$ the Mn magnetic structure was inferred indirectly from synchrotron x-ray diffraction data which show the presence of an incommensurate lattice modulation with \qlMn\space varying between 0.72 and 0.77 below \TN = 39~K.\cite{kimura2} By analogy with the Tb compound, \qmMn = 0.36...0.385 in DyMnO$_{3}$. No information on the Dy magnetic structure has been reported yet. The Dy ordering, however, causes much more significant anomalies in the dielectric constant $\epsilon$ and the electric polarization $P$ than the Tb ordering in TbMnO$_{3}$.\cite{goto}
Therefore, the coupling between the RE and the Mn appears to be more important in DyMnO$_{3}$ than in the Tb compound.
 
In this Communication we report on synchrotron x-ray diffraction and resonant magnetic scattering experiments on single crystal DyMnO$_{3}$ carried out at temperatures between 4 and 40~K. Below \TNDy = 5~K, the Dy magnetic moments order in a commensurate structure with propagation vector \qmDy \bb, \qmDy = 0.5. Simultaneous with the Dy magnetic ordering, an incommensurate lattice modulation with \qlDy = 0.905 evolves while the original Mn induced modulation is suppressed and the value of \qlMn shifts from 0.78 to 0.81. This behavior points to a strong interference of Dy and Mn induced structural distortions in DyMnO$_3$ besides a magnetic coupling between the Dy and Mn magnetic moments. The findings are related to reported effects of the Dy ordering on the static electric properties.

\begin{figure}[bt!]
\begin{center} 
\includegraphics[scale=0.7]{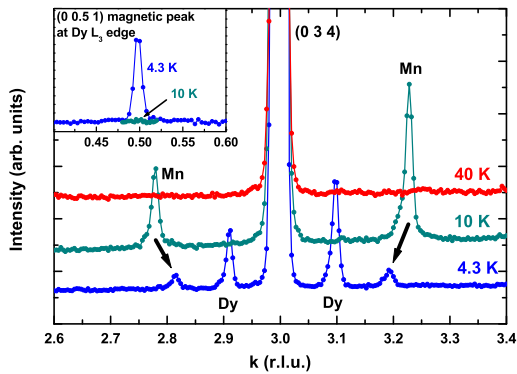} 
\caption{(Color online) $k$-scans along $(0k4)$ at 40~K (top line), 10~K (center), and 4.3~K (bottom), $E = 12.4$~keV. Data are offset for clarity. A similar set of curves was observed for $l = 3$ where the Dy peaks are about a factor of ten larger relative to the Mn ones. Inset: $k$-scan along $(0k1)$ around $k = 0.5$ measured at 4.3 and 10~K with $E = 7.794$~keV.}
\label{Fig1}
\end{center}
\end{figure}

A DyMnO$_{3}$ single crystal was obtained by recrystalizing a ceramic rod under oxygen atmosphere using an optical floating zone furnace. The crystal develops nice cleavage planes perpendicular to the $b$ axis of the orthorhombic structure.
The crystal quality was checked by x-ray Laue diffraction and conventional laboratory four-circle diffractometry confirming that the crystal shows the orthorhombically distorted perovskite structure with space group $Pbnm$ ($a= 5.27$~\AA, $b=5.80$~\AA, $c=7.36$~\AA).

Synchrotron x-ray diffraction measurements were conducted on the new 7~T multipole wiggler beamline MAGS, operated by the HMI at the synchrotron source BESSY in Berlin. This beamline provides a focused beam in the energy range 4 to 30~keV, monochromatized by a Si(111) double crystal monochromator ($\sim$2~eV energy resolution), with a photon flux of the order of 10$^{12}$ photons/s at 10~keV. The experimental endstation consists of a six-circle diffractometer which can be equipped optionally with a three-circle analyzer for linear polarization analysis. A displex-type cryostat provided a base temperature slightly above 4~K. All temperature dependences were measured on heating in order to avoid hysteresis effects. Temperature stability was $\pm 0.2$~K.

The measurements were carried out in vertical scattering four-circle geometry at a photon energy of 12.398~keV ($\lambda =1$~\AA) and close to the Dy L$_3$-edge at 7.790~keV. For the latter measurements the energy calibration was checked with a thin Co foil as standard. At 12.4~keV the transversal and longitudinal resolution were $6\times10^{-4}$\inA and $4\times10^{-3}$\inA, respectively. For some of the experiments around 7.79~keV the (006) reflection of a HOPG crystal (0.35\degg\space mosaic) was used in the polarization analyzer, with an analyzer Bragg angle 2$\theta$ = 90.7\degg, close to 90\degg. The scattered photons were detected by a scintillation counter. Higher order x-rays were suppressed by appropriate settings of the beamline mirror tilt angles and the detector discriminator windows. The sample with dimensions 4 mm diameter, 5 mm length was mounted to a Cu sample holder with the $bc$ plane perpendicular to the scattering plane, so that virtually all reflections of type $(0kl)$ could be reached. To avoid sample heating by x-ray irradiation - observed with the full beam intensity at temperatures below 5 K - the power of the beam was reduced by one order of magnitude with appropriate absorber foils.

\begin{figure}[bt!]
\begin{center} 
\includegraphics[scale=0.7]{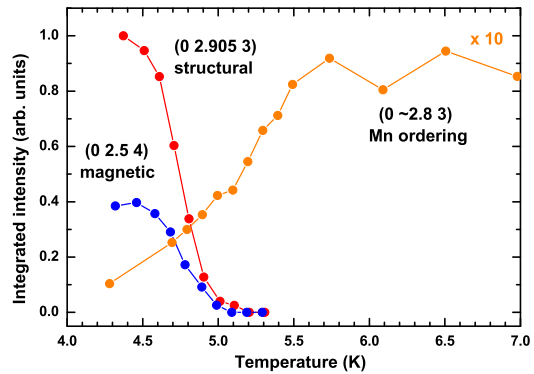} 
\caption{(Color online) Temperature dependence of the integrated intensities of the Dy induced  $(0~2.905~3)$ and $(0~2.5~4)$ superstructure reflections together with that of the Mn induced $(0~2.8~3)$ around the magnetic transition of Dy. To match scales, the latter was multiplied with a factor of 10.}
\label{Fig2}
\end{center}
\end{figure}

First the structural modulations were investigated at temperatures between 4 and 40~K using 12.398~keV photon energy. Broad $k$-scans were carried out along the $(0 k l)$ direction for $l = 3$ and 4. FIG.~1 shows the results obtained for $l = 4$. 
At 40~K (top line), above \TN, no superstructure reflections are present. At 10 K (center), below \TNMn\space but still above \TNDy, the superstructure reflections (0~$2+$\qlMn~4) and (0~$4-$\qlMn~4) induced by the Mn ordering are clearly visible as a satellite pair (0~3$\pm$\dmn~4), with \dmn\space $= 1 - $\qlMn\space = 0.22, consistent with previous x-ray diffraction results.\cite{kimura2} Finally, at 4.3~K (bottom), these Mn induced reflections have shifted to \dmn\space = 0.19 and decreased significantly in intensity. An additional pair of superstructure reflections is found at \ddy\space = 0.095. We assign these reflections to the lattice modulation induced by the Dy magnetic ordering, i.e., \qlDy\space = 0.905 ($= 1-$\ddy), in analogy with TbMnO$_{3}$, where \qlTb $\sim 6/7$ has been observed.\cite{kenz,kajimoto}
From the observed Lorentzian broadening of the Mn (10~K) and Dy (4.3~K) reflections with respect to the main Bragg reflections
we infer values for the coherence lengths associated with the corresponding structural modulations of $\sim$450~\AA\space and $\sim$500~\AA\space, respectively, along the $b$ axis.

\begin{figure}[bt!]
\begin{center} 
\includegraphics[scale=0.7]{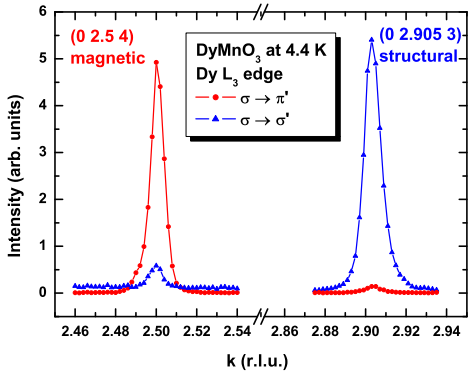} 
\caption{(Color online) Linear polarization analysis of (left) the $(0~2.5~4)$ and (right) the $(0~2.905~3)$ superstructure reflections. The intensity variation as a function of the polarization analyzer configuration shows that $(0~2.5~4)$ is of magnetic and $(0~2.905~3)$ of structural origin.}
\label{Fig3}
\end{center}
\end{figure}

To look for the Dy \emph{magnetic} superstructure reflections, the x-ray energy was tuned to 7.794~keV, slightly above the Dy-L$_3$ absorption edge. The initial measurements were carried out without polarization analysis. The inset in FIG.~1 shows a $k$-scan along $(0k1)$ around $k = 0.5$ measured at 4.3~K. Surprisingly, we find a single reflection at the commensurate position $k = 0.5$ instead of the expected satellite pair at $k = 0.455$ and 0.545 (\qlDy/2 and $1-$\qlDy/2). From the observed linewidth we infer
a corresponding coherence length $\gtrsim$500~\AA. A survey of reciprocal space yielded a number of additional reflections of type $(0~n+0.5~l)$, where $(0~2.5~4)$ turned out to be the most intense. All these reflections vanish above 5~K. FIG.~2 shows the temperature dependence of the integrated intensities of the $(0~2.905~3)$ and $(0~2.5~4)$ superstructure reflections together with that of the Mn induced reflection at $(0~2.8~3)$. On heating, the two former disappear simultaneously at the Dy ordering temperature \TNDy = 5~K, suggesting that they have the same origin.

In order to identify the nature of the two Dy-induced reflections, in a next step we employed linear polarization analysis. Here, the diffracted beam is reflected by an analyzer crystal with a deflection angle of $\sim$90\degg. It can be rotated so that the reflection occurs either in the original scattering plane ($\sigma\rightarrow\sigma^{\prime}$ configuration) or at 90\degg\space to this plane ($\sigma\rightarrow\pi^{\prime}$). FIG.~3 shows the dependence of the intensities of both the $(0~2.5~4)$ and the $(0~2.905~3)$ superstructure reflections on the analyzer configuration. Structural scattering from a lattice modulation does not change the original 95\% horizontal polarization of the wiggler beam, so that we expect very little intensity (zero for perfect linear polarization) in a corresponding signal measured in $\sigma\rightarrow\pi^{\prime}$ configuration. This obviously is the case for the (0~2.905~3) reflection. Magnetic scattering, on the other hand, strongly favors the $\sigma\rightarrow\pi^{\prime}$ channel if the ordered magnetic moments have components parallel to the scattering plane. This clearly is the case for the $(0~2.5~4)$ reflection. The residual $(0~2.5~4)$ signal observed in the $\sigma\rightarrow\sigma^{\prime}$ configuration is 
presumably caused by a significant spin component perpendicular to the scattering plane. 
A further confirmation of the magnetic origin of the $(0~2.5~4)$ reflection is the energy dependence of the peak intensity close to the Dy-L$_3$ edge, shown in FIG.~4. It shows a resonance enhancement by a factor of at least twenty, while the structural reflections (not shown) exhibit a reduction of intensity due to the increased absorption at the edge. We can therefore assign a magnetic propagation vector \qmDy \bb, \qmDy = 0.5 to the Dy magnetic ordering in DyMnO$_{3}$. The same propagation vector has been observed for the Ho ordering in orthorhombic HoMnO$_{3}$ below 6 K.\cite{munoz}

\begin{figure}[bt!]
\begin{center} 
\includegraphics[scale=0.7]{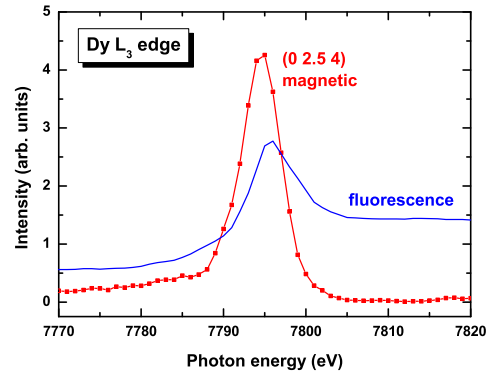} 
\caption{(Color online) X-ray energy dependence of the intensity of the $(0~2.5~4)$ reflection close to the Dy-L$_3$ edge, measured with polarization analyzer in $\sigma\rightarrow\pi^{\prime}$ configuration, together with the fluorescence yield.}
\label{Fig4}
\end{center}
\end{figure}

One important observation of the present study is the suppression of the original Mn induced lattice modulation by the magnetic ordering of Dy. This is illustrated in more detail by the evolution of the Dy induced $(0~2.905~3)$ and the Mn induced $(0~2.8~3)$ reflections close to \TNDy shown in FIG.~5. The incommensurate position of the $(0~2.905~3)$ reflection does not vary significantly ($\delta k < 0.002$) below 5~K, in apparent contrast to the behavior of the Mn superlattice reflection which exhibits a complex behavior. Above 7 K it consists of a single component somewhat broader than the $(0~2.905~3)$. At lower temperatures it shifts slightly towards higher $k$ values and develops a shoulder at $k = 2.81$ already about 0.5~K above \TNDy, i.e., \emph{precursor} to the Dy magnetic ordering. Simultaneously the total integrated intensity of the reflection starts to drop (see FIG.~2), the component at $k = 2.81$ gains relative intensity at the expense of the $k = 2.78$ peak, and at base temperature the latter has practically disappeared. This behavior is presumably caused by a strong magnetic coupling between the Dy and Mn moments via exchange fields. Owing to its large magnetic moment of up to 10~$\mu_{B}$ the onset of commensurate correlations between the Dy magnetic moments apparently leads to a suppression of the original incommensurate spiral Mn spin structure accompanied by a slight shift of the wave vector \qmMn\space of the residual modulation towards the value of \qmDy. 

\begin{figure}[bt!]
\begin{center} 
\includegraphics[scale=0.7]{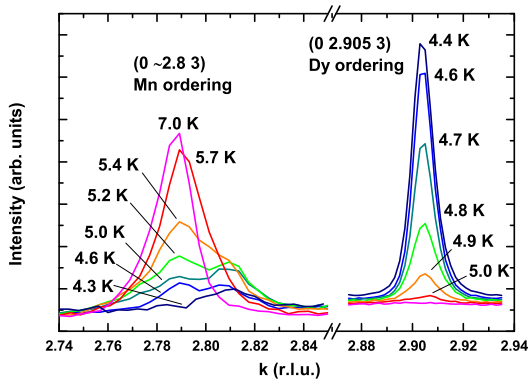} 
\caption{(Color online) Temperature evolution of the Dy induced (0~2.905~3) and the Mn induced (0~2.8~3) around the magnetic transition of Dy, measured at the Dy L$_3$-edge.}
\label{Fig5}
\end{center}
\end{figure}

A more intriguing result of the present study is the formation of an additional incommensurate lattice modulation at the transition to the commensurate magnetic order of the Dy in DyMnO$_{3}$. One possible explanation would be the formation of a new modulated spin structure of the Mn with propagation vector $q_{~m}^{\prime \rm Mn} = 0.45$~\bb\space induced by the exchange fields from the ordered Dy ions. However, this would imply the coexistence of two Mn modulations with different propagation vectors below \TNDy, because the original modulation is not completely suppressed. This scenario therefore appears unlikely. A second explanation has to invoke a strong interference between Mn and Dy induced lattice distortions. Conventionally one would expect the Mn and Dy magnetic orderings to induce their own lattice modulations which would not couple assuming they involve non-interacting displacements of Mn and Dy atoms.
However, our measurements show that in DyMnO$_{3}$ lattice modulations with different periodicity induced by the Mn and Dy magnetic orderings do not simply superpose. From quadratic magnetoelastic coupling the Dy magnetic ordering would be expected to produce a commensurate lattice modulation with propagation vector 1~\bb. This modulation apparently interferes with the existing modulation 0.78~\bb. The result appears to be a compromise, with a final dominant modulation 0.905~\bb and the residual of the original Mn induced modulation shifted to 0.81~\bb. Interestingly, the two sets of superstructure reflections observed at 4.3~K exactly obey the matching \dmn = 2\ddy (where $\delta = 1-q_{l}$, see above). This suggest a strong magneto-elastic coupling between the two different types of modulations, presumably mediated by displacements of oxygen atoms.\cite{kimura2,aliouane:020102} For comparison, in TbMnO$_{3}$ the periodicities of the Mn and Tb magnetic structures, \qmMn $\sim 0.277$ and \qmTb $\sim 3/7$, lead to two lattice modulations with \dmn $\sim$ 3$\delta^{Tb}$ matching in a similar way as in the Dy compound without the need for any shift of $q_{l}$ values. Accordingly, in TbMnO$_{3}$ only a marginal shift of the Mn induced superstructure reflections is observed below \TNTb.\cite{kenz} 

It is interesting to relate these observations to the reported effects of the Tb and Dy magnetic orderings on the static electric properties of TbMnO$_{3}$ and DyMnO$_{3}$, respectively. In the former, Tb ordering causes only small anomalies in the dielectric constant $\epsilon$ and the electric polarization $P$. In contrast, the Dy ordering causes much more significant anomalies in these properties. Especially the pronounced kink at 6~K in the temperature dependence of \pc\space with a steep drop of the value of \pc\space between 6 and 5~K\cite{goto} apparently is closely related to the suppression and shift of the spiral Mn spin structure precursor to the Dy ordering. A similar jump around 6~K is observed in the value of $\epsilon_{a}$. The magneto-elastic interaction between Dy and Mn may support the suppression of the spiral Mn spin structure in this temperature range. In more general terms, our measurements demonstrate the significant coupling between lattice distortions arising from Mn and RE magnetic ordering which should not be ignored from theoretical models that attempt to predict the multiferroic properties of these perovskites as a function of magnetic field.

In conclusion, we have shown that in DyMnO$_{3}$, below \TNDy = 5~K, the Dy magnetic moments order with a commensurate propagation vector 0.5~\bb. The magnetic coupling between the Dy and Mn magnetic moments leads to a suppression of the incommensurate spiral Mn spin structure and a shift of the associated lattice modulation wave vector from 0.78~\bb to 0.81~\bb. This is the origin of the observed changes of the static electric properties of DyMnO$_{3}$ around \TNDy. The ordering of the Dy magnetic moments is accompanied by an incommensurate lattice modulation 0.905~\bb, pointing to a strong interference of Dy and Mn induced structural distortions in DyMnO$_{3}$, presumably mediated by displacements of oxygen atoms.

\emph{Acknowledgments} - Construction of the beamline MAGS has been founded by the BMBF via the HGF-Vernetzungsfonds under contracts No.~01SF0005 and 01SF0006. We thank the BESSY staff for technical support and S. Landsgesell for assistance in the sample preparation.

\end{document}